\documentclass[9pt,twocolumn,twoside,superscriptaddress]{revtex4-1}

\RequirePackage[utf8]{inputenc}
\RequirePackage[english]{babel}
\RequirePackage{amsmath,amsfonts,amssymb}
\RequirePackage{mathpazo} 
\RequirePackage[scaled]{helvet}
\RequirePackage[T1]{fontenc}
\RequirePackage{url}
\RequirePackage[colorlinks=true, allcolors=blue]{hyperref}

\usepackage{graphicx,xcolor}

\usepackage{color}
\definecolor{mycyan}{cmyk}{1,0,0,0.12} 

\begin{document}

\author{Carlos Abellan}
\affiliation{ICFO - Institut de Ciencies Fotoniques, The Barcelona Institute of Science and Technology, 08860 Castelldefels (Barcelona), Spain}
\affiliation{Corresponding author: carlos.abellan@icfo.es}
\author{Waldimar Amaya}
\affiliation{ICFO - Institut de Ciencies Fotoniques, The Barcelona Institute of Science and Technology, 08860 Castelldefels (Barcelona), Spain}
\author{David Domenech}
\affiliation{VLC Photonics S.L. Cami de Vera s/n, Edificio 9B}
\author{Pascual Mu\~noz}
\affiliation{VLC Photonics S.L. Cami de Vera s/n, Edificio 9B}
\affiliation{ITEAM Research Institute, Universitat Polit\`ecnica de Valencia, Spain}
\author{Jose Capmany}
\affiliation{VLC Photonics S.L. Cami de Vera s/n, Edificio 9B}
\affiliation{ITEAM Research Institute, Universitat Polit\`ecnica de Valencia, Spain}
\author{Stefano Longhi}
\affiliation{Dipartimento di Fisica and Istituto di Fotonica e Nanotecnologie del CNR, Politecnico di Milano, Milan (Italy)}
\author{Morgan W. Mitchell}
\affiliation{ICFO - Institut de Ciencies Fotoniques, The Barcelona Institute of Science and Technology, 08860 Castelldefels (Barcelona), Spain}
\affiliation{ICREA - Instituci\'o Catalana de Recerca i Estudis Avancats, 08015 Barcelona, Spain}
\author{Valerio Pruneri}
\affiliation{ICFO - Institut de Ciencies Fotoniques, The Barcelona Institute of Science and Technology, 08860 Castelldefels (Barcelona), Spain}
\affiliation{ICREA - Instituci\'o Catalana de Recerca i Estudis Avancats, 08015 Barcelona, Spain}

\title{A quantum entropy source on an InP photonic integrated circuit for random number generation}

\begin{abstract}
Random number generators are essential to ensure performance in information technologies, including cryptography, stochastic simulations and massive data processing. The quality of random numbers ultimately determines the security and privacy that can be achieved, while the speed at which they can be generated poses limits to the utilisation of the available resources. In this work we propose and demonstrate a quantum entropy source for random number generation on an indium phosphide photonic integrated circuit made possible by a new design using two-laser interference and heterodyne detection. The resulting device offers high-speed operation with unprecedented security guarantees and reduced form factor. It is also compatible with complementary metal-oxide semiconductor technology, opening the path to its integration in computation and communication electronic cards, which is particularly relevant for the intensive migration of information processing and storage tasks from local premises to cloud data centres.
\\\\
\url{http://dx.doi.org/10.1364/OPTICA.3.000989}
\end{abstract}

\maketitle

\section{Introduction}
Random numbers (RNs) are essential to a wide range of applications, including secure communications to protect the transmission and storage of confidential data \cite{Shannon:1949dt}, massive data processing \cite{Brin:1998vm}, and stochastic simulations \cite{Mascagni:2014fe, Click:2011gm} for stock market predictions, decision making in engineering processes and Monte Carlo calculations of physical, chemical, nuclear and biological events. Pseudo- RNs can be generated through computational algorithms while true RNs can only be generated through physical processes \cite{VonNeumann:1951va}. Quantum mechanical processes are the best guarantee for offering high-performance without compromising security and privacy. So far, several quantum entropy sources (QES) have been proposed for quantum random number generation (QRNG), including single photon splitting \cite{Rarity:1994uu}, homodyne detection of the vacuum field \cite{Gabriel:2010gba}, and phase diffusion (PD) in semiconductor lasers \cite{Qi:2010vf, Jofre:2011wv}. To date, PD-QRNGs have achieved the highest bit rates \cite{Abellan:2014tv, Yuan:2014coa, Nie:2015gs} , up to 68 Gb/s \cite{Nie:2015gs}, and passed severe random tests \cite{Abellan:vv}. However, so far, PD-QRNGs have been realized with discrete optical components, often leading to devices of large size. 

\begin{figure*}[!t]
\centering
{\includegraphics[width=0.9\linewidth]{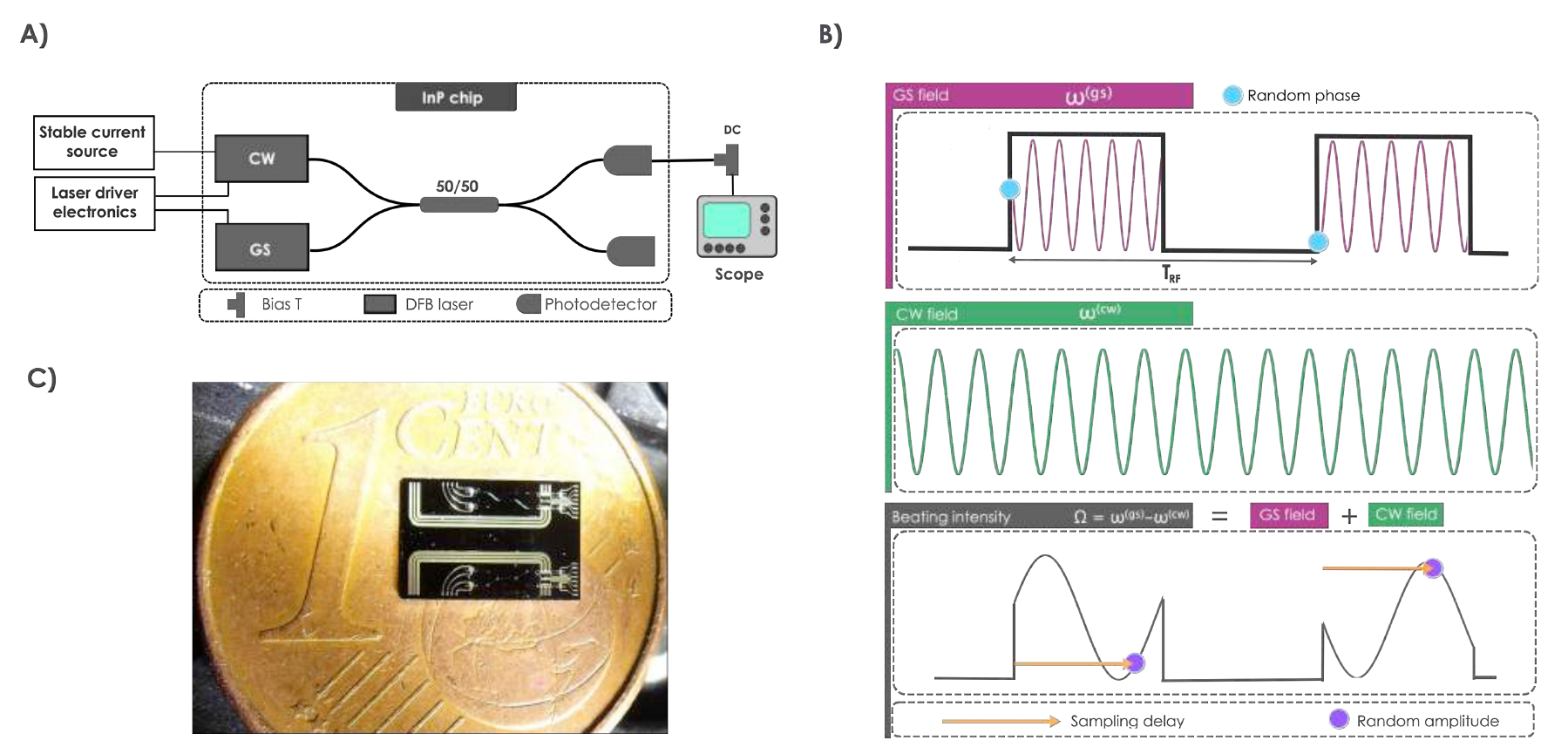}}
\caption{(a) Schematic of the quantum random number generator photonic integrated circuit (QRNG-PIC) based on two-laser interference. The two distributed feedback (DFB) lasers are biased with a current driver, one of them operating in continuous wave (CW) while the other one is periodically gain switched (GS) using an external RF generator. The temperature of the entire chip is controlled through a Peltier element while that of the area including one of the lasers is locally changed by a stable current source. The outputs from the two lasers are combined and interfered in a 2x2 multimode interference (MMI) coupler and two 40 GHz photodiodes (PDs) are placed at the output of the coupler. The detected signal is sent to a fast oscilloscope. (b) Principle of operation: optical pulses from a GS laser interfere with a CW laser generating an interference modulation whose frequency is equal to the difference of the two lasers' frequencies. The random phase of the GS laser pulse produces a random phase of the interference oscillation that can be properly sampled into a random amplitude. In this way, after digitization, one can extract one sample per GS pulse. (c) Microscope image of the PIC on a 1 Euro cent background. Two QRNG-PICs are printed on each chip. }
\label{fig:fig1}
\end{figure*}

Photonic integrated circuit (PIC) technology \cite{Heck2012, Smit:2014je} is a key ingredient for building scalable optical devices \cite{Anonymous:6VYIFizk}. The telecommunication industry is a clear example, and already accounts for commercial products such as semiconductor lasers, 100 GHz photo detectors, and high-bandwidth optical interconnects and transceivers \cite{Alduino:10}. Recently, the quantum optics community is making rapid progress in leveraging PIC technology, offering the possibility to build scalable quantum optics experiments. In the field of quantum computation, PIC technology in combination with additional bulk elements, such as lasers, is allowing for the development of novel experiments otherwise impossible using tabletop components. Some examples include quantum simulation \cite{Tillmann:2013jv} and quantum enhanced sensing \cite{Matthews:2009gi}. Quantum key distribution (QKD) functionalities have been also integrated using Indium Phosphide technology \cite{Sibson:15} and a monolithically integrated QRNG, composed of a light-emitting-diode (LED) and a single-photon-avalanche-photodetector (SPAD), has been recently demonstrated at 1 Mb/s using Silicon (Si) photonics technology \cite{Khanmohammadi:2015cx}. Si photonics is a promising candidate for building scalable optical applications due to the compatibility with the microelectronics industry. However, the impossibility of creating a Si laser source poses serious limitations to the level of integration and performance of the PIC. 

In this work, we show a fully integrated quantum entropy source for random number generation on an InP platform (QES-PIC) using standard fabrication techniques only. The device is made possible by a new design using two-laser interference and heterodyne detection, allowing QRNG rates in the Gb/s regime. We observe high interference visibility during long execution runs as well as superior temperature stability when compared to the bulk implementation of the same scheme. Also, using the Lang-Kobayashi rate equations model, we study in detail the dynamics of the two integrated lasers. We find conditions for operating the two-lasers with a negligible coupling effect, and provide an accurate description and modeling of the strong thermal chirp observed in the InP distributed feedback (DFB) lasers. 


\section{Experiment}
\label{sec:experiment}
As illustrated in Fig. \ref{fig:fig1}(A), we introduce a new QES scheme that combines two DFB lasers on the same chip. The first laser is operated in gain switching (GS) mode while the second one in continuous wave (CW) mode. By modulating continuously the GS laser from below to above threshold, optical pulses with nearly identical waveforms and completely randomized phases are generated. Then, by beating the GS and CW (the local oscillator) lasers through a multimode interference (MMI) coupler, an intensity oscillation forms with a beating frequency equal to the difference of the two lasers' frequencies, which can be detected by a photodetector (PD), see Fig. \ref{fig:fig1}(B). Being $i^{(cw)}$ and $i^{(gs)}$ the intensities from the CW and the GS lasers respectively, we can write the total intensity at the output of the MMI (see supplementary material) as 
\begin{equation}
i_T(t) = i_S(t) + 2i_P(t)\cos \Big( \int_0^t d\xi \Omega_C(\xi) + \Delta\phi\Big)
\end{equation}
where $i_S (t)\equiv i^{(cw)}+i^{(gs)}$ is the sum of the intensities from the two lasers, $i_P (t)\equiv(i^{(cw)} i^{(gs)})^{1/2}$ the geometric mean, $\Delta\phi=\phi^{(cw)}-\phi^{(gs)}$ the phase difference between the two lasers fields, and $\Omega_C(t) = \Omega - \beta(t)$ the frequency detuning as a function of time. We introduced $\beta(t) = \beta_0 t$ phenomenologically to account for frequency chirp arising from fast thermal effects in the directly-modulated laser \cite{Zadok:1998bf}. Here, $\Omega$ represents the initial frequency detuning between the two lasers. As illustrated in Fig. \ref{fig:fig1}(B), the resulting signal corresponds to a train of pulses in which the amplitude of each pulse oscillates at $\int_0^t d\xi \Omega_c(\xi)$ with a random phase $\Delta\phi$ (for simplicity, $\beta_0=0$ in the illustration). Finally, after the MMI coupler, a photodetector converts the optical signal into the electrical domain and random numbers are obtained by taking one sample per period.

A microscope image of the two laser QES-PIC is shown in Fig. \ref{fig:fig1}(C). The chip was placed on top of a Peltier controller and its temperature was maintained at 25$^\circ$ with variations below 0.1$^\circ$. The first DFB laser, with a bias of 10 mA, was operated in GS mode by superimposing a 100 MHz modulation from an Anritsu MP1800A pulse generator through a bias tee port. We chose this relatively low modulation frequency to capture properly the dynamics of the interference pattern within the GS pulse. However, modulation frequencies up to 2 GHz are within immediate reach, allowing for 10s of Gb/s raw generation rates using current analog-to-digital conversion technologies, these being only limited by the stabilization time of the build-up dynamics of the laser intensity. The CW laser was operated by applying a constant 30 mA current. The beating signal was detected by an on-chip 40 GHz photodetector and digitized with a 20 GHz and 50 GSa/s real-time scope (Digital Phosphor Oscilloscope Tektronix DPO72004C), providing a temporal resolution of 20 ps to analyse the beat-note. 

The central frequency of the two lasers could be independently tuned by injecting a constant current from a stable source Keithley 2401 through a metallic contact on the grating structure. The heating effect changed the average refractive index of the grating and consequently the Bragg condition, thus the operating frequency (wavelength) of each laser. By tuning these currents, the detuning frequency between the two lasers could be reduced and brought within the detection bandwidth. 

\section{Results}

\subsection{Modelling}
When the frequencies of the two lasers are tuned closed to each other, back reflections from the MMI coupler can give rise to phase (frequency) locking effects. This phenomenon can be explained on the basis of the general mechanism of Adler's synchronisation of two coupled nonlinear oscillators \cite{Adler:1946db, Aronson, aronson2} and modeled by the Lang-Kobayashi rate equation analysis for two mutually-coupled semiconductor lasers \cite{Lang:1980kw, Yanchuk:2004cr, Erzgraber:2005gd}. Indicating by $\kappa$ the effective coupling rate between the two laser cavities and by $\Omega$ the frequency detuning between the two bare longitudinal modes of the uncoupled cavities, neglecting delay effects it is known in simple Adler's theory of synchronisation that frequency locking occurs for $|\Omega|<2\kappa$  \cite{Adler:1946db}. A more accurate analysis can be gained from rate equation analysis. We model the coupled laser system with rate equations for the normalised complex slowly varying envelope of the optical fields $E_{1,2}$ and the normalised inversions $N_{1,2}$ using standard Lang- Kobayashi rate equations \cite{Lang:1980kw, Yanchuk:2004cr, Erzgraber:2005gd}, which in dimensionless form read  \cite{Erzgraber:2005gd}
\begin{align}
\label{Eq:2}\dot{E}_1 &= \gamma(1+\mathrm{i}\alpha)N_1E_1 + \kappa\exp({\mathrm{i}\psi}) E_2(t-\tau_d) + \sqrt{R_1}\xi_1(t)\\
\notag \dot{E}_2 &= \gamma(1+\mathrm{i}\alpha)N_2E_2 + \kappa\exp({\mathrm{i}\psi}) E_1(t-\tau_d) \\
\label{Eq:3}&\hspace{3cm}+ \mathrm{i}[\Omega+\beta(t)]E_2 \sqrt{R_2}\xi_2(t)\\
\label{Eq:4}\tau \dot{N}_{1} &= P_{1} - N_{1} - (1+2N_{1})|E_{1}|^2\\
\label{Eq:5}\tau \dot{N}_{2} &= P_{2} - N_{2} - (1+2N_{2})|E_{2}|^2
\end{align}

In Eqs. (\ref{Eq:2}-\ref{Eq:5}), $\alpha$ is the linewidth enhancement factor, $\gamma$ is the photon decay rate in the two laser cavities, $\tau$ is the carrier lifetime, $\kappa$ is the coupling rate due to spurious optical feedback, $\tau_d$ and $\psi$ are the time and phase delays of optical feedback, $P_{1,2}(t)$ are the normalized pump parameters of the two lasers, and $\Omega$ is the difference between the oscillating frequencies of the two uncoupled lasers. The normalized pump parameter $P$ is given by $P=G_N N_0 (x-1)/2$, where $G_N$ is the differential gain, $N_0$ is the current density at threshold, and $x=J/J_{th}$ is the actual pumping current density normalized to its threshold value. Spontaneous emission is modeled by Langevin forces describing a Gaussian white noise process $\xi(t)$ with zero mean and correlations given by $\langle \xi_R(t) \xi_R (t')\rangle = \langle \xi_l(t) \xi_l (t')\rangle=\delta(t-t')$, and $\langle \xi_R (t) \xi_I (t')\rangle=0$ \cite{Torcini2006}. For the laser operated with a carrier density not too far from its threshold value, the noise term becomes additive with a variance $R=R_{sp}$, where $R_{sp}$ is the spontaneous emission variance \cite{Torcini2006}.

To simulate the experimental results, we assume that the first laser is pumped with a constant current $(P_1= \bar{P_1})$, whereas the second laser is periodically GS from below to above threshold with a current pulse $P_2(t)$. In the simulations, the normalized pump parameter for the second laser in each modulation cycle is assumed to be of the form 
\begin{equation}
P_2(t) = \bar{P_2}\Big\{ -\frac{1}{2} + \frac{3}{2}\exp[-(t/\Delta\tau)^{2M}]\Big\},
\end{equation}
where $\bar{P_2}$ is the peak of normalized pump current, $\Delta\tau$ the current pulse duration and $M$ the super-Gaussian parameter. 

\begin{figure}[!b]
\centering
{\includegraphics[width=0.9\linewidth]{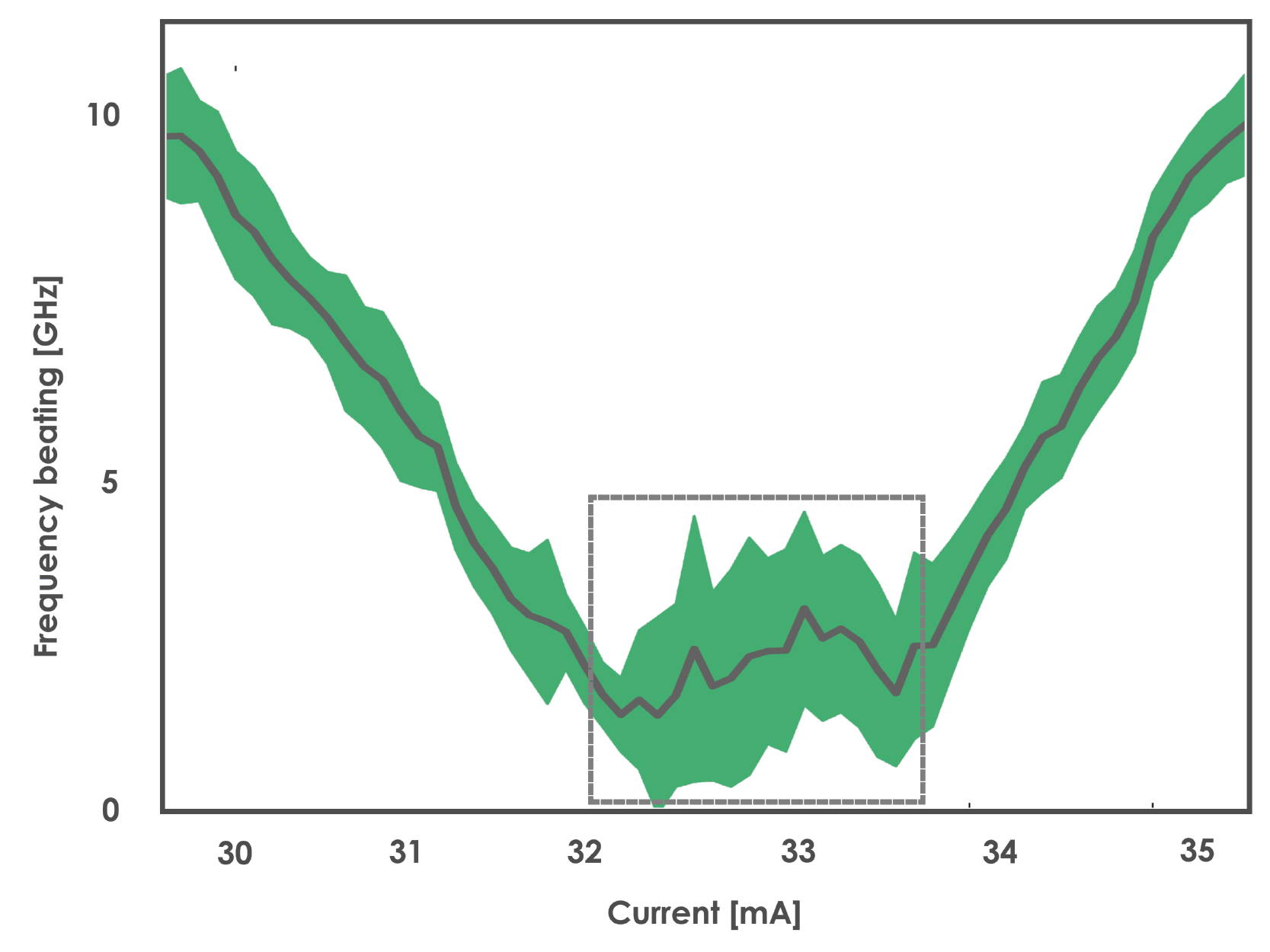}}
\caption{Beat-note frequency at the output of the MMI measured by sweeping one of the integrated lasers while keeping the other one constant in the low-loss QRNG-PIC. The beat- note frequency can be continuously tuned by current control for large detuning frequencies, whereas for small detuning frequencies phase (frequency) locking may occur (grey square), leading eventually to disappearance of the oscillation. }
\label{fig:fig2}
\end{figure}

\begin{figure*}[!t]
\centering
{\includegraphics[width=0.9\linewidth]{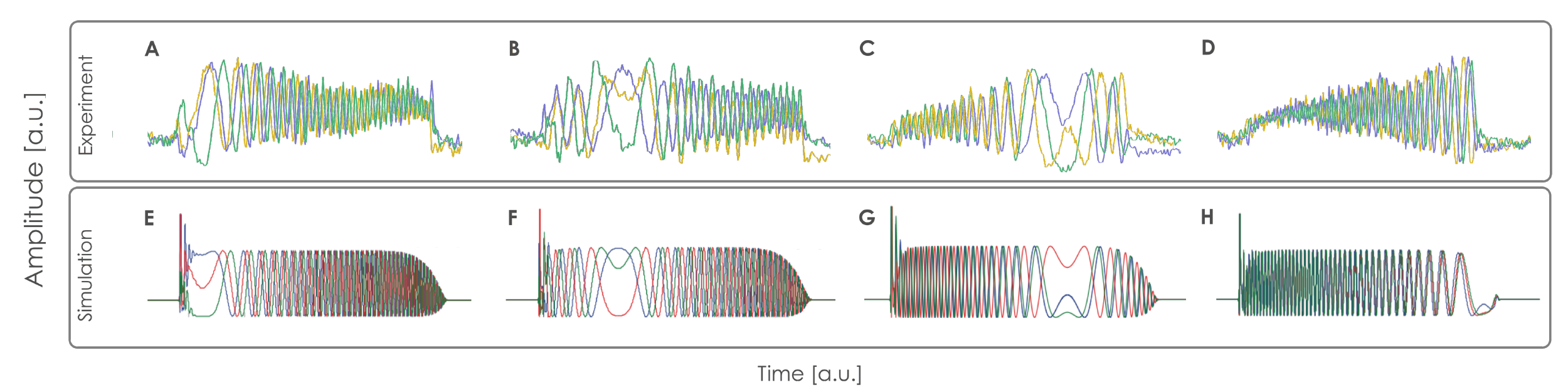}}
\caption{Temporal dynamics of the beating between the two lasers forming the high-loss QRNG-PIC and comparison with numerical results. (A-D) Experimental data with different temperature settings (currents). Chirp due to thermal effects and attenuation of beating amplitude due to the bandwidth limit of the detection electronics are evident. (E-H) Numerical results with initial detuning frequencies set to fit the experimental observations in (a). }
\label{fig:fig3}
\end{figure*}

\subsection{Numerical simulation and experimental parameters}
In the experiments we used two PICs: a high waveguide propagation-loss PIC that ensures absence of significant back reflection from the CW into the GS laser and another one with a similar structure but lower loss. In the lower loss PIC, back-reflections at the MMI interface eventually induce phase-locking effects, preventing the generation of random oscillations. The coupling constant $\kappa$ can be estimated from the locking region of the two lasers in the low-loss chip, see Fig. \ref{fig:fig2}, and is given by $\kappa = 5$ ns$^{-1}$. With such a high coupling constant, the Langevin forces entering in Eqs. (\ref{Eq:2}-\ref{Eq:5}) are too weak and therefore phase locking is observed in spite of spontaneous emission noise. 

To obtain the QRNG functionality, it is therefore mandatory to reduce the feedback arising from spurious reflections at the MMI coupler. This goal was simply achieved by increasing the optical losses of the bus waveguides ($\sim 15 $ dB/mm). As a result, the coupling rate $\kappa$ between the two laser cavities is reduced by about $\sim 30$ dB from the previously discussed low-loss PIC. For this high-loss PIC, the coupling, if any, is very weak, and thus phase randomization due to quantum noise prevails over phase locking, as shown in Fig. \ref{fig:fig3}. Parameter values used in the simulation are shown in Table I. We use the instantaneous coupling approximation \cite{Yanchuk:2004cr}, i.e. the delay $\tau_d$ can be neglected when solving Eqs. (\ref{Eq:2}-\ref{Eq:3}), since the time delay and coupling rate of feedback satisfy the constraint $\tau_d \kappa < 1/\sqrt{1+\alpha^2}$. On the other hand, the phase delay $\psi$ is difficult to estimate owing to its strong sensitivity to length changes over one wavelength; as a matter of fact it can be considered as an independent variable \cite{Aronson, aronson2}. Nevertheless, for parameter values that apply to our experimental conditions, a change of the phase delay, e.g. from $\psi=0$ to $\psi=\pi/2$, does not introduce substantial changes in the beating dynamics.

\begin{table}[htbp]
\centering
\caption{\bf The value of the chirp rate $\beta_0$ has been chosen to qualitatively fit the experimental results and is consistent with the data reported in \cite{Zadok:1998bf}. The delay time $\tau_d$ is determined by the optical path between the laser output facet and the MMI coupler. }
\begin{tabular}{ccc}
\hline
Parameter & Symbol & Value \\
\hline
Linewidth enhancement factor & $\alpha$ & $2$ \\
Carrier lifetime & $\tau$ & $1$ ns \\
Photon decay rate & $\gamma$ & $150$ ns$^{-1}$\\
Feedback delay time & $\tau_d$ & $20$ ps \\
Normalised pump parameter & $\bar{P_1} = \bar{P_2}$  & 8 \\
Current pulse duration & $\Delta \tau$ & $5$ ns \\
Super-Gaussian parameter & $M$ & 5 \\
Chirp rate & $\beta_0$ & $2\pi \times 1$ MHz/ns \\
Spontaneous emission rate & $R_{sp}$ & $2\times 10^{-4}$ ps$^{-1}$\\
\hline
\end{tabular}
  \label{tab:sim_params}
\end{table}

\subsection{Measuring the beating dynamics}
For both the high- and low-loss PICs, the optical pulses of the GS laser were strongly chirped due to thermal effects, yielding a frequency-varying oscillation of the beating pattern, as depicted in Fig. \ref{fig:fig3}. As a result, a nearly-zero-detuning (NZD) region was observed within the optical pulses when the chirped frequency of the GS laser coincides with the stable frequency of the CW laser. The position of the NZD region depends on the initial frequency separation between the GS and the CW emission lines. When both lasers were initially close (far) in frequency, the NZD region occurred at the beginning (end) of the pulse, see Fig. \ref{fig:fig3}. In the high (15 dB/cm) waveguide loss PIC, the interference amplitude within the NZD region changed from pulse to pulse, a clear signature that phase noise dominated. Instead, in the low loss PIC (2 dB/cm), back reflection from the CW into the GS laser was not negligible and phase locking between the two lasers was observed. In this case, the interference amplitude in the NZD region did not appreciably change from pulse to pulse. 

In the experiment, the NZD region was tuned at the end of the pulse (see Fig. \ref{fig:fig3}(D)) maximasing the detuning frequency between the two lasers so as to reduce residual phase locking effects, if any. 

\subsection{PIC stability and performance}
\begin{figure}[!t]
\centering
{\includegraphics[width=\linewidth]{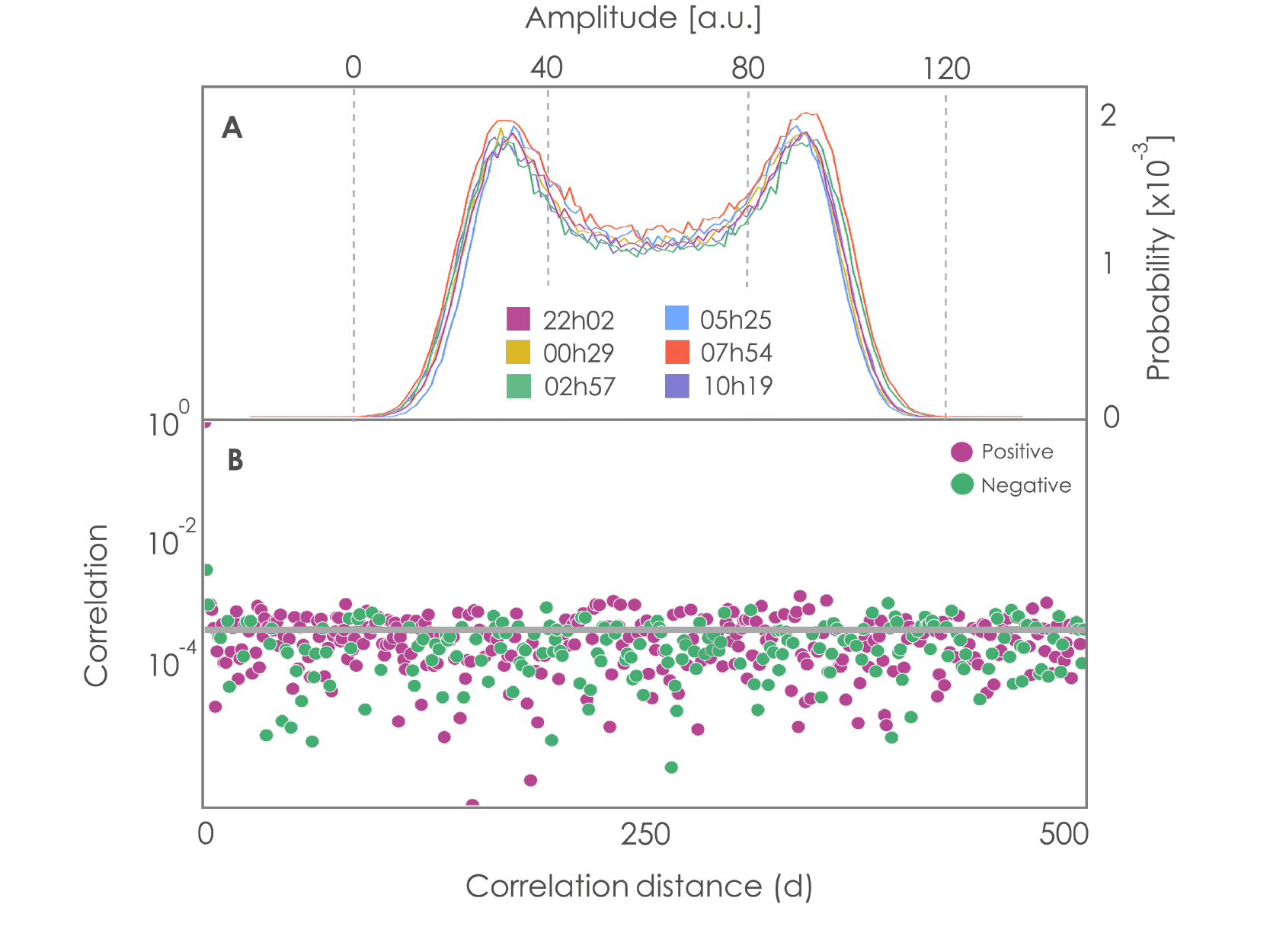}}
\caption{Statistics on the output of the QES-PIC. (a) Histograms on 6 sets of $200.000$ samples each taken during 14 hrs, confirming stable operation of the QES-PIC device. (b) Autocorrelation function for $10^7$ random samples taken with a 20 GHz scope and 50 GSa/s. Magenta (green) circles correspond to positive (negative) correlation coefficients. }
\label{fig:fig4}
\end{figure}

From a practical point of view, long-term stability of the scheme is a critical aspect. As we are interfering signals from two independent lasers, intrinsic phase noise and temperature drifts can severely affect the performance. In Fig. \ref{fig:fig4}(A), we plot the histogram for 6 data sets with $200.000$ samples each. The digitised signal was distributed according to the arcsine probability distribution function because of the initial random phase \cite{Abellan:2014tv, Yuan:2014coa} and the digitisation frequency asynchronous with respect to the beating (detuning) frequency. High stability was observed between acquisitions taken during 14 hrs., confirming the robustness of the two-laser scheme QES-PIC. Instead, a similar implementation with discrete (bulk) components suffered from slow temperature drifts (see supplementary material for details on the bulk implementation and corresponding experimental results). The higher stability of the PIC over the bulk design was mainly associated to the fact that the two lasers are closely located in a region with uniform temperature.

In Fig. \ref{fig:fig4}(B), we show the autocorrelation function $\Gamma_x(k)\equiv \langle x_i x_{i+k}\rangle - \langle x_i\rangle^2$ of a sequence of $n=10^7$ samples up to a delay of $500$ samples. For such sequence length, the statistical uncertainty due to finite size effects is $3.16\times 10^{-4}$. Except for the $d=1$ coefficient, which is significantly larger than the statistical noise sensitivity, all the other coefficients fall within the statistical noise level and pass the D'Agostino-Person's normality test with a $p$-value of $0.18$. We attribute the larger correlation at $d=1$ to limitations in the direct modulation of the DFB LD in the experiment, leading to residual photons in the cavity from pulse to pulse. Samples from the higher loss PIC were acquired using a $50$ GSa/s resolution and $20$ GHz bandwidth real time scope, followed by a 30 dB RF amplifier. The amplifier introduced noise at several frequency bands, so we employed a $30$ MHz high pass-band digital filter to remove low-frequency components. In addition, we want to assess the quality of the QES, so we want to analyze the correlation of the beat signal only. For doing so, we assume the noise is independent of our signal and calculate $\Gamma_x = \Gamma_y - \Gamma_n$, where $\Gamma_y$ corresponds to the autocorrelation of samples taken within the GS pulse, and $\Gamma_n$ to samples taken outside the GS pulse. 

\section{Discussion}
The production of application-ready random numbers from the QES requires a randomness extraction stage \cite{Frauchiger:2013tf}. In real QRNG devices, untrusted noise degrades (corrupts) the purity of the randomness associated to quantum processes. The application of proper randomness extractors allows to eliminate corruption of the quantum signal \cite{Frauchiger:2013tf}. Random extraction requires (i) the estimation of the amount of available min-entropy from the QES, taking into account electronic noise, memory effects and digitization noise \cite{Mitchell:2015fj}, and (ii) an appropriate hashing of the data after digitization. Thus, a full QRNG solution that includes both the QES and the electronics (including the digitizer) is required before any meaningful entropy estimate can be derived. However, the statistical data reported in Fig.~\ref{fig:fig4} (correlation and distribution) already confirms the high randomness quality, as it is comparable to the raw data obtained with previously demonstrated bulk architectures, in which large entropy rates have been reported employing different digitization strategies in complete PD-QRNG solutions \cite{Abellan:vv,Mitchell:2015fj}. With respect to the post-processing algorithm, field-programmable-gate-arrays (FPGAs) can be used for real-time randomness extraction above 1 Gb/s \cite{Martin:2015jg} for high-performance applications, while for lower-end applications, such as consumer electronics, the central processing unit (CPU) can sustain up to several Mb/s \cite{Sanguinetti:2014cc}. 

\section{Conclusion}
We have presented an ultra-fast quantum entropy source on a photonic integrated circuit (QES-PIC) for use in quantum random number generation (QRNG). The resulting device shows high performance, including bit rate, degree of randomness (low correlation values) and stability, in a miniaturised geometry that also integrates the receiver.  Considering its small footprint and the possibility for hybrid integration with CMOS electronics, the proposed QES-PIC has the potential to become a future functionality in computer and communication cards, especially for cryptography and stochastic simulations. The high level of miniaturisation may even make the integration of the QES-PIC device in smartphones and tablets possible.

\section*{Funding Information}
European Research Council project AQUMET, European Union Project QUIC (Grant Agreement No. 641122), Spanish MINECO under the Severo Ochoa programme (Grant No. SEV-2015-0522), project EPEC (Grant No. FIS2014- 62181-EXP), project Qu-CARD (SRTC1400C002844XV0), Catalan AGAUR 2014 SGR Grants No. 1295 and No. 1623, the European Regional Development Fund (FEDER) Grant No. TEC2013-46168- R, and by Fundaci\'o Privada CELLEX. 

\section*{Acknowledgments}
We thank M. Jofre and M. Curty for stimulating discussions on the optical design and R. Ba\~nos for laboratory assistance. \\\\


\noindent See Supplement 1 for supporting content.

%
%
%
%
%
%


\end{document}